\documentclass{sig-alternate}
\usepackage{algorithm}
\usepackage{url}
\newdef{ex}{Example}
\newdef{definition}{Definition}
\newdef{theorem}{Theorem}
\usepackage{algorithmic}
\usepackage{multirow}
\begin{document}

\title{A DNF Blocking Scheme Learner for Heterogeneous Datasets}

\date{3rd February, 2014}

\maketitle
\begin{abstract}
Entity Resolution concerns identifying co-referent entity pa-\\irs across datasets. A typical workflow comprises two steps. In the first step, a blocking method uses a one-many function called a blocking scheme to map entities to blocks. In the second step, entities sharing a block are paired and compared. Current DNF blocking scheme learners (DNF-BSLs) apply only to structurally homogeneous tables. We present an unsupervised algorithmic pipeline for learning DNF blocking schemes on RDF graph datasets, as well as structurally heterogeneous tables. Previous DNF-BSLs are admitted as special cases. We evaluate the pipeline on six real-world dataset pairs. Unsupervised results are shown to be competitive with supervised and semi-supervised baselines. To the best of our knowledge, this is the first unsupervised DNF-BSL that admits RDF graphs and structurally heterogeneous tables as inputs.
\end{abstract}

\category{I.2.6}{Artificial Intelligence}{Learning}
\category{H.2.8}{Database Management}{Database applications}[Data Mining]

\terms{Algorithms, Experimentation, Performance}

\keywords{Heterogeneity, Entity Resolution, Unsupervised Learning}

\section{Introduction}
\emph{Entity Resolution} (ER) is the identification of co-referent entities across datasets. Different communities refer to it as \emph{instance matching}, \emph{record linkage} and the \emph{merge-purge problem} \cite{recordlinkagesurvey, ERsurvey}. Scalability indicates a two-step solution \cite{recordlinkagesurvey}. The first step, \emph{blocking}, mitigates brute-force pairwise comparisons on all entities by clustering entities into blocks and then comparing pairs of entities only within blocks \cite{christensurvey}. Blocking results in the selection of a small subset of pairs, called the \emph{candidate} set, which is input to a second step to determine co-reference using a sophisticated similarity function. We exclusively address blocking in this work.

Blocking methods use a \emph{blocking scheme} to assign entities to blocks. Recently, Disjunctive Normal Form (DNF) Blocking Scheme Learners (BSLs) were proposed to learn DNF blocking schemes using supervised \cite{bilenkoblocking, knoblockblocking}, semi-supervised \cite{semisup} or unsupervised machine learning techniques \cite{mayankblocking}.

DNF-BSLs operate in an expressive hypothesis space and have shown excellent empirical performance \cite{bilenkoblocking}. Despite their advantages, current DNF-BSLs assume that input dat-\\asets are \emph{tabular} and have the \emph{same schemas}. The latter assumption is often denoted as \emph{structural homogeneity} \cite{recordlinkagesurvey}. 

The assumption of tabular structural homogeneity restricts application of DNF-BSLs to other data models. Recent growth of graph datasets and Linked Open Data (LOD) \cite{bernerslinked} motivates the development of a DNF-BSL for the RDF\footnote{Resource Description Framework} data model. These graphs are often published by independent sources and are \emph{heterogeneous} \cite{silk}. 

In this paper, we present a \emph{generic algorithmic pipeline} for learning DNF blocking schemes on pairs of RDF datasets. The formalism of DNF blocking schemes relies on the existence of a \emph{schema}. RDF datasets on LOD may not have accompanying schemas \cite{bernerslinked}. Instead of relying on possibly unavailable metadata, we build a \emph{dynamic schema} using the properties in the RDF dataset. The RDF dataset can then be \emph{logically} represented as a \emph{property table}, which may be populated at run-time. Previously, property tables were defined as \emph{physical} data structures used in the implementation of \emph{triplestores} \cite{propertytable}. Using a logical property table representation admits application of a DNF-BSL to RDF datasets.      

As a special case, the pipeline also admits \emph{structurally heterogeneous} tabular inputs. That is, the pipeline can be applied to tabular datasets with \emph{different} schemas. Thus, previous DNF-BSLs become special cases of the pipeline, since they learn schemes on tables with the same schemas. As a second special case, the pipeline accommodates \emph{RDF-tabular heterogeneity}, with one input, RDF, and the other, tabular.  RDF-tabular heterogneity applies when linking datasets between LOD and the relational Deep Web \cite{deepweb, bernerslinked}. 

Much existing ER research bypasses the issue of structural heterogeneity by assuming that \emph{schema matching} is physically undertaken prior to executing an ER algorithm \cite{recordlinkagesurvey}. Schema matching itself is an active, challenging research problem \cite{schemamatching,schematough}. Instead of relying on the unrealistic assumption of perfect schema reconciliation, the pipeline admits possibly noisy schema mappings from an existing schema matcher, and \emph{directly} uses the mappings to learn a DNF blocking scheme. If a matcher is unavailable, the pipeline allows for a \emph{fall-back} option.   

We show an \emph{unsupervised instantiation} of the generic pipe-\\line by strategically employing an existing instance-based schema matcher called \emph{Dumas} \cite{dumas}. In addition to schema mappings, Dumas also generates noisy duplicates, which we \emph{recycle} and pipe to a \emph{new} DNF-BSL. The new DNF-BSL uses only two parameters, which may be robustly tuned for good empirical performance across multiple domains. The BSL is also shown to have a strong \emph{theoretical} guarantee. 

We evaluate the full unsupervised pipeline on six real-world dataset pairs against two extended baselines, one of which is \emph{supervised} \cite{bilenkoblocking, mayankblocking}. Our system performance is found to be competitive, despite training samples not being manually provided. Given that unsupervised methods exist for the \emph{second} ER step in both the Semantic Web and relational communities \cite{ERsurvey, recordlinkagesurvey}, this first unsupervised heterogeneous DNF-BSL \emph{enables}, in principle, fully unsupervised ER in \emph{both} communities.    

Section \ref{relatedwork} describes some related work in the area, followed by preliminaries in Section \ref{prelims}. Section \ref{pipeline} describes the generic pipeline, followed by an unsupervised instantiation in Section \ref{unsup}. Section \ref{experiments} describes the experiments, with results discussed in Section \ref{results}. The paper concludes in Section \ref{conclusion}. 

\section{Related Work}\label{relatedwork}
ER was comprehensively surveyed by Elmagarmid et al. \cite{recordlinkagesurvey}, with a generic approach represented by  \emph{Swoosh} \cite{swoosh}. Separately, blocking has witnessed much specific research, with Christen surveying blocking methods \cite{christensurvey}.  Bilenko et al. \cite{bilenkoblocking}, and Michelson and Knoblock \cite{knoblockblocking} independently proposed supervised DNF-BSLs in 2006. Since then, a semi-supervised adaptation of the BSL proposed by Bilenko et al. has been published \cite{bilenkoblocking, semisup}, as well as an unsupervised system \cite{mayankblocking}. The four systems assume \emph{structural homogeneity}. We discuss their core principles in Section \ref{leds}.

Heterogeneous blocking may be performed without learning a DNF scheme. One example is Locality Sensitive Hashing (LSH), employed by the \emph{Harra} system, for instance \cite{harra}. LSH is promising but applies only to specific distance measures, like Jaccard and cosine.  The \emph{Typifier} system by Ma et al. is another recent example that relies on \emph{type inferencing} and was designed for Web data published without semantic types \cite{typifier}. In contrast, DNF-BSLs can be applied \emph{generally}, with multiple studies showing strong empirical performance \cite{knoblockblocking, bilenkoblocking, semisup, mayankblocking}.  Finally, although related, \emph{clustering} is technically treated separately from blocking in the literature \cite{christensurvey}.
 
In the Semantic Web, ER is simultaneously known as \emph{instance matching} and \emph{link discovery}, and has been surveyed by Ferraram et al. \cite{ERsurvey}. Existing works restrict inputs to RDF. Also, most techniques in the Semantic Web do not \emph{learn} schemes, but instead present graph-based blocking \emph{methods}, a good example being Silk \cite{silk}. 

Finally, the framework in this paper also relies on \emph{schema mapping}. Schema mapping is an active research area, with a good survey provided by Bellahsene et al. \cite{schemamatching}. Gal notes that it is a difficult problem \cite{schematough}. 

Schema matchers may return 1:1 or n:m mappings (or even 1:n and n:1). An \emph{instance-based} schema matcher relies on data instances to perform schema matching \cite{schemamatching}. A good example is \emph{Dumas} \cite{dumas}, which relies on an inexpensive \emph{duplicates generator} to perform unsupervised schema matching \cite{dumas}. We describe Dumas in Section \ref{unsup}.

The \emph{property table} representation used in this paper is a \emph{physically} implemented data structure in the Jena triplestore API\footnote{\url{https://jena.apache.org/}} \cite{propertytable}. In this paper, it is used as a \emph{logical} data structure. We note that the concept of logically representing one data model as another has precedent. In particular, the literature abounds with proposed methods on how to integrate relational databases (RDB) with the Semantic Web. Sahoo et al. extensively surveyed this topic, called \emph{RDB2RDF} \cite{rdb2rdf}. A use-case is the \emph{Ultrawrap} architecture, which utilizes RDB2RDF to enable real-time Ontology-based Data Access or OBDA \cite{ultrawrap}. We effectively tackle the \emph{inverse} problem by translating an RDF graph \emph{to} a logical property table. This is the first application to devise such an inverse translation for heterogeneous ER.

\section{Preliminaries}\label{prelims}
\begin{figure*}[t]
\centering
\epsfig{file=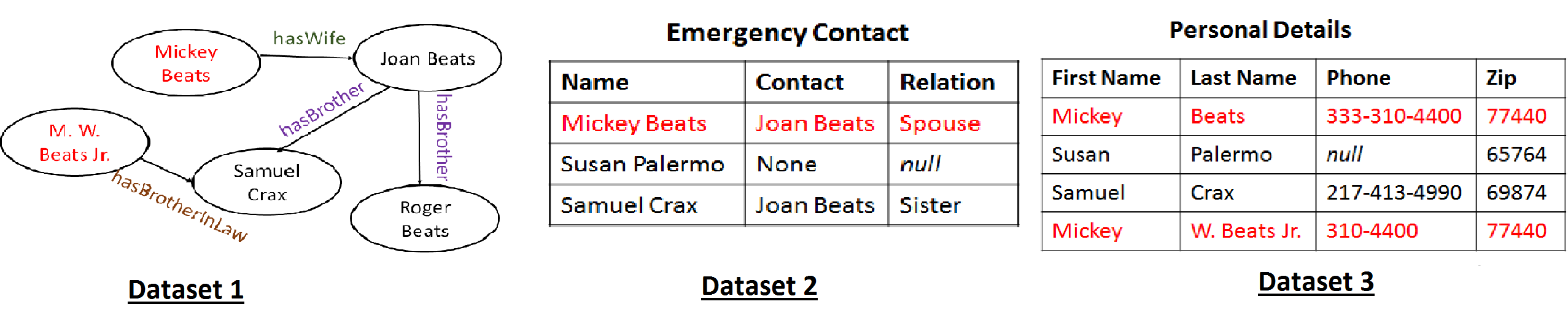, height=1.2in, width=6.0in}
\caption{Three example datasets exhibiting various kinds of heterogeneity}
\label{datasetsfig}
\end{figure*}

We present definitions and examples to place the remainder of the work in context.
Consider a pair of datasets $R_1$ and $R_2$. Each dataset individually conforms to either the RDF or tabular data model. An \emph{RDF} dataset may be visualized as a \emph{directed graph} or equivalently, as a set of \emph{triples}. A triple is a 3-tuple of the form \emph{(subject, property\footnote{Alternatively called \emph{predicate}; we uniformly use \emph{property}.}, object)}. A \emph{tabular} dataset conforms to a \emph{tabular schema}, which is the table name followed by a set of \emph{fields}. The dataset \emph{instance} is a set of \emph{tuples}, with each tuple comprising \emph{field values}.  
\begin{ex}
Dataset 1 (in Figure \ref{datasetsfig}) is an RDF dataset visualized as a directed graph $G=(V,E)$, and can be equivalently represented as a set of $|E|$ triples. For example, \emph{(Mickey Beats, hasWife, Joan Beats)} would be one such triple in the triples representation.
Datasets 2 and 3 are tabular dataset examples, with the former having schema \emph{Emergency Contact(Name, Contact, Relation)}. The first tuple of Dataset 2 has field values \emph{Mickey Beats}, \emph{Joan Beats} and \emph{Spouse} respectively. The keyword \emph{null} is reserved. 
\end{ex}

According to the RDF specification\footnote{\url{http://www.w3.org/RDF/}}, subjects and properties must necessarily be \emph{Uniform Resource Identifiers} (URIs), while an object node may either be a URI or a literal. URI elements in RDF files typically have associated \emph{names} (or \emph{labels}), obtained through dereference. For ease of exposition, we henceforth refer to every URI element in an RDF file by its associated name. Note also, that in the most general case, RDF datasets do not have to conform to any schema. This is why they are commonly visualized as \emph{semi}-structured datasets, and not as tables. In Section \ref{proptable}, we show how to dynamically build a \emph{property schema} and logically represent RDF as a tabular data structure.  

An entity is defined as a \emph{semantically} distinct \emph{subject} node in an RDF dataset, or as a (semantically distinct) \emph{tuple} in a tabular dataset. The entity referring to \emph{Mickey Beats} is shown in red in all datasets in Figure \ref{datasetsfig}. \emph{Entity Resolution} is the process of resolving \emph{semantically equivalent} (but possibly \emph{syntactically different}) entities. As earlier described, ER is traditionally conducted on tables with the same schemas, or on \emph{one}\footnote{Commonly just called \emph{deduplication}, if this is the case \cite{recordlinkagesurvey}.} dataset. An example would be identifying that the two highlighted tuples in Dataset 3 are duplicates. 

In the Semantic Web, ER is operationalized by connecting two equivalent entities with an \emph{owl:sameAs}\footnote{\url{http://www.w3.org/TR/owl-ref/}} property edge. For example, the two nodes referring to Mickey Beats in Dataset 1 should be connected using an \emph{owl:sameAs} edge. Easy operationalizing of ER (and more generally, \emph{link specification} \cite{silk}) explains in part the recent interest in ER in the Semantic Web \cite{ERsurvey}. In the relational setting, ER is traditionally operationalized through joins or mediated schemas. It is less evident how to operationalize ER across RDF-tabular inputs, such as linking Datasets 1 and 2. We return to this issue in Section \ref{proptable}.

To introduce the \emph{current} notion of DNF blocking schemes, tabular structural homogeneity is assumed for the remainder of this section. In later sections, we generalize the concepts. 

The most basic elements of a blocking scheme are \emph{indexing functions} $h_i(x_t)$ \cite{bilenkoblocking}. An indexing function accepts a field value from a tuple as input and returns a set $Y$ that contains 0 or more \emph{blocking key values} (BKVs). A BKV identifies a block in which the tuple is placed. Intuitively, one may think of a block as a hash bucket, except that blocking is one-many while hashing is typically many-one \cite{christensurvey}. For example, if $Y$ contains multiple BKVs, a tuple is placed in multiple blocks.  
\begin{definition}\label{if}
An \emph{indexing function} $h_i : Dom(h_i) \rightarrow U^{*}$ takes as input a field value $x_t$ from some tuple $t$ and returns a set $Y$ that contains 0 or more \emph{Blocking Key Values} (BKVs) from the set of all possible BKVs $U^{*}$.
\end{definition}

The domain $Dom(h_i)$ is usually just the \emph{string} datatype. The range is a set of BKVs that the tuple is assigned to. Each BKV is represented by a string identifier. 
\begin{ex} 
An example of an indexing function is \emph{Tokens}. When applied to the \emph{Last Name} field value of the fourth tuple in Dataset 3, the output set $Y$ is $\{W., Beats, Jr.\}$.
\end{ex}

This leads to the notion of a \emph{general blocking predicate} (GBP). Intuitively, a GBP $p_i(x_{t_1},x_{t_2})$ takes as input field values from two tuples, $t_1$ and $t_2$, and uses the $i^{th}$ indexing function to obtain BKV sets $Y_1$ and $Y_2$ for the two arguments. The predicate is satisfied \emph{iff} $Y_1$ and $Y_2$ share elements, or equivalently, if  $t_1$ and $t_2$ have a block in common.
\begin{definition}\label{gbp}
A \emph{general blocking predicate} $p_i: Dom(h_i) \times Dom(h_i) \rightarrow \{True, False\}$ takes as input field values
$x_{t_1}$ and $x_{t_2}$  from two tuples, $t_1$ and $t_2$, and returns \emph{True} if $h_i(x_{t_1}) \cap h_i(x_{t_2}) \neq \Phi$, and returns \emph{False} otherwise.
\end{definition}

Each GBP is always associated with an indexing function. 
\begin{ex}
Consider the GBP \emph{ContainsCommonToken}, associated with the previously introduced \emph{Tokens}. Suppose it was input the \emph{Last Name} field values from the first and fourth tuples in Dataset 3. Since these field values have a token (\emph{Beats}) in common, the GBP returns \emph{True}.
\end{ex}

A \emph{specific blocking predicate} (SBP) explicitly pairs a GBP to a specific field.
\begin{definition}\label{sbp}
A \emph{specific blocking predicate} is a pair $(p_i,f)$ where $p_i$ is a general blocking predicate and $f$ is a field. A specific blocking predicate takes two tuples $t_1$ and $t_2$ as arguments and applies $p_i$ to the appropriate field values $f_1$ and $f_2$ from both tuples. A tuple pair is said to be \emph{covered} if the specific blocking predicate returns \emph{True} for that pair. 
\end{definition}
 
Previous DNF research assumed that all available GBPs can be applied to \emph{all} fields of the relation \cite{mayankblocking,bilenkoblocking, semisup, knoblockblocking} . Hence, given a relation $R$ with $m$ fields in its schema\footnote{Structural homogeneity implies exactly one input schema, even if there are multiple relational instances.}, and $s$ GBPs, the number of SBPs is exactly $ms$. Finally, a \emph{DNF blocking scheme} is defined as: 
\begin{definition}\label{dnfbs}
A \emph{DNF blocking scheme} $f_P$ is a \emph{positive} propositional formula constructed in \emph{Disjunctive Normal Fo-\\rm}\footnote{A disjunction of \emph{terms}, where each term is a conjunction of literals.} (DNF), using a given set $H$ of SBPs as the set of \emph{atoms}. Additionally, if each \emph{term} is constrained to comprise at most one atom, the blocking scheme is referred to as \emph{disjunctive}.
\end{definition} 

SBPs cannot be negated, since the DNF scheme is a positive formula. A tuple pair is said to be \emph{covered} if the blocking scheme returns \emph{True} for that pair. Intuitively, this means that the two constituent tuples \emph{share} a block. In practice, both duplicate and non-duplicate tuple pairs can end up getting covered, since blocking is just a pre-processing step.  
\begin{ex}
Consider the disjunctive scheme \emph{(ContainsC-\\ommonToken, Last Name) $\vee$ (SameFirstDigit, Zip)}, applied on Dataset 3. While the two tuples referring to Mickey Beats would share a block (with the BKV \emph{Beats}), the non-duplicate tuples referring to Susan and Samuel would \emph{also} share a block (with the BKV \emph{6}). Note also that the first and fourth tuples share \emph{more} than one block, since they also have BKV \emph{7} in common. 
\end{ex}

Given a blocking scheme, a \emph{blocking method} would need to map tuples to blocks \emph{efficiently}. Per Definition \ref{dnfbs}, a blocking scheme takes a tuple pair as input. In practice, linear-time hash-based techniques are usually applied.
\begin{ex}
To efficiently apply the blocking scheme in the previous example on each \emph{individual} tuple, tokens from the field value corresponding to field \emph{Last Name} are extracted, along with the first character from the field value of the \emph{Zip} field, to obtain the tuple's set of BKVs. For example, applied to the first tuple of Dataset 3, the BKV set $\{Beats, 7\}$ is extracted. An index is maintained, with the BKVs as keys and tuple pointers as values. With $n$ tuples, traditional blocking computes the blocks in time $O(n)$ \cite{christensurvey}. 
\end{ex}

Let the set of generated blocks be $\Pi$. $\Pi$ contains sets of the form $B_v$, where $B_v$ is the \emph{block} referred to by the BKV $v$. The \emph{candidate set} of pairs $\Gamma$ is given below:
\begin{equation}\label{gamma}
\Gamma=\bigcup_{B_v\in\Pi} \{(r,s)\}, \forall r,s \in B_v | r \in R_1, s \in R_2   
\end{equation}  

$\Gamma$ is precisely the set input to the \emph{second} step of ER, which classifies each pair as a duplicate, non-duplicate or probable duplicate \cite{datamatching}. Blocking should produce a small $\Gamma$ but with high coverage and density of duplicates. Metrics quantifying these properties are defined in Section \ref{experiments}.

Finally, \emph{schema mapping} is utilized in the paper. The \emph{formal} definition of a mapping is quite technical; the survey by Bellahsene et al. provides a full treatment \cite{schemamatching}. In this paper, an intuitive understanding of the mapping as a \emph{pair} of \emph{field-sets} suffices. For example, (\{Name\},\{First Name, Last Name\}) is a 1:n mapping between Datasets 2 and 3. More generally, mappings may be of cardinality n:m. The simplest case is a 1:1 mapping, with singleton components. 

\section{The generic pipeline}\label{pipeline}
\begin{figure*}[t]
\centering
\epsfig{file=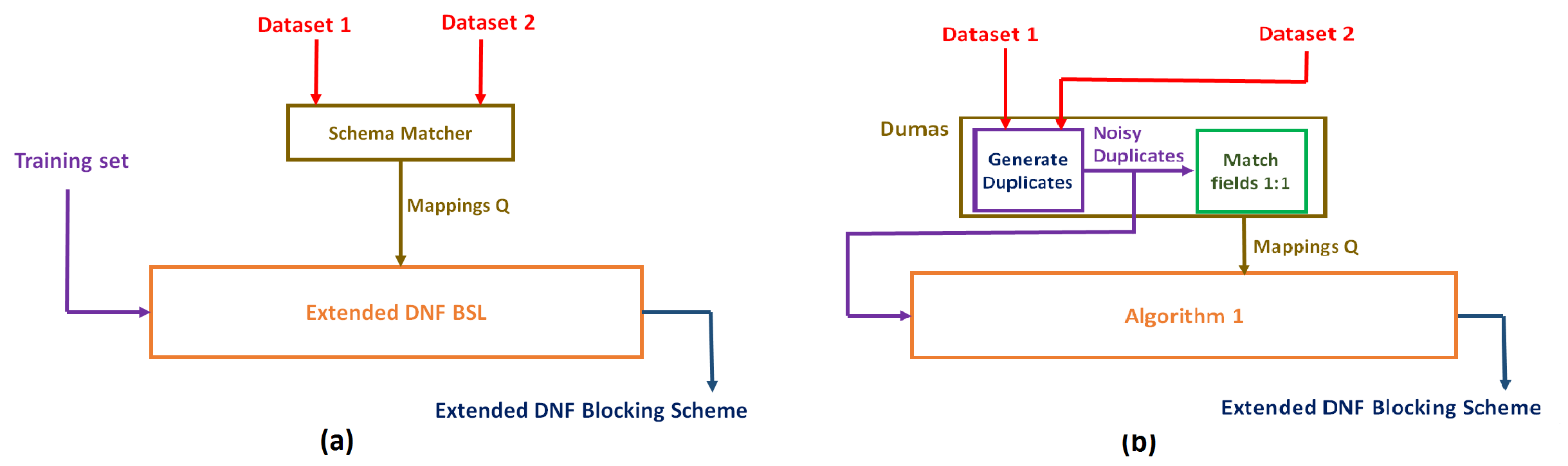, height=1.5in, width=7.20in}
\caption{(a) shows the generic pipeline for learning a DNF blocking scheme on heterogeneous datasets, with (b) showing an \emph{unsupervised instantiation}}
\label{pipelinefig}
\end{figure*}
Figure \ref{pipelinefig} (a) shows a schematic of the \emph{generic} pipeline. Two heterogeneous datasets are initially provided as input, with either dataset being RDF or tabular. If the dataset is RDF, we logically represent it as a \emph{property table}. We describe the details and advantages of this \emph{tabular} data structure in Section \ref{propertytable}. The key point to note is that the \emph{schema matching} module takes two \emph{tables} as input, regardless of the data model, and outputs a set of schema mappings $Q$. An \emph{extended} DNF-BSL accepts $Q$ and also a training set of duplicates and non-duplicates as input, and learns an extended DNF blocking scheme. The DNF-BSL and the blocking scheme need to be extended because tables are now structurally heterogeneous, and the Section \ref{prelims} formalism does not natively apply. In Section \ref{dnf}, the formalism is extended to admit structural heterogeneity.

Figure \ref{pipelinefig} (b) shows an \emph{unsupervised instantiation} of the generic pipeline. We describe the details  in Section \ref{unsup}.  

\subsection{Property Table Representation}\label{propertytable}
Despite their formal description as sets of triples, RDF files are often \emph{physically} stored in triplestores as sparse property tables \cite{propertytable}. In this paper, we adapt this table instead as a \emph{logical} tabular representation of RDF. To enable the logical construction on RDF dataset R, define the \emph{property schema} $\{subject\} \cup \{p|\exists (s,p,o), (s,p,o) \in R\}$. In essence, we flatten the graph by assigning each distinct \emph{property} (or edge label) a corresponding field in this schema, along with an extra \emph{subject} field. Every distinct subject in the triples-set has exactly one corresponding tuple in the table.
\begin{figure}\label{proptable}
\centering
\epsfig{file=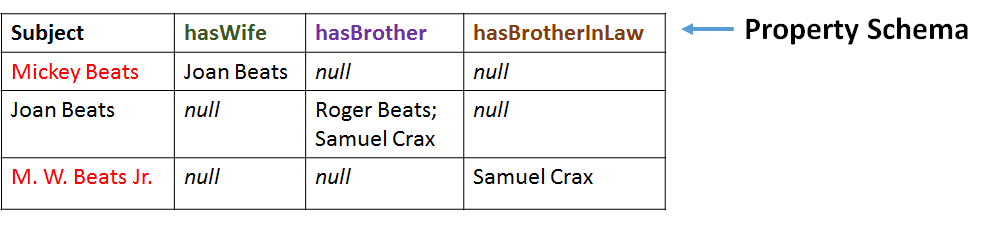, height=0.8in, width=3.3in}
\caption{Property Table representation of Dataset 1 in Figure \ref{datasetsfig}}
\label{proptablefig}
\end{figure}

For example, Figure \ref{proptablefig}  is the property table representation of Dataset 1 in Figure \ref{datasetsfig}. If a subject does not have a corresponding object value for a given property, the reserved keyword \emph{null} is entered. If a subject has \emph{multiple} object values for a property, the values are concatenated using a reserved delimiter (\emph{;} in Figure \ref{proptablefig}). Technically, field values now have \emph{set} semantics, with \emph{null} representing the empty set. Also, the original set of triples can be losslessly reconstructed from the property table (and vice versa), making it an \emph{information-preserving} logical representation. Although intuitively straightforward, we detail these lossless conversions in the Appendix.

The \emph{physical} property table was proposed to eliminate expensive \emph{property-property self-joins} that occur frequently in SPARQL\footnote{\url{http://www.w3.org/TR/rdf-sparql-query/}} queries. For ER, the logical data structure is useful because it allows for a \emph{dynamic schema} that is resolvable at \emph{run-time}. Triplestores like Jena allow updating and querying of RDF \emph{graphs}, despite the underneath tabular representation \cite{propertytable}. If the RDF dataset is already stored in such a triplestore, it would not have to be moved prior to ER. This gives the property table a \emph{systems-level} advantage.

More importantly, having a schema for an RDF dataset means that we can invoke a suitable schema matcher in the generalized pipeline. As we subsequently show, the extended DNF-BSL in the pipeline requires the datasets to have (possibly different) schemas\footnote{Even non-DNF blocking typically assumes this \cite{typifier}.}. Finally, representing RDF as a table allows us to address \emph{RDF-tabular heterogeneity}, by reducing it to tabular structural heterogeneity. Traditional ER operations become well-defined for RDF-tabular inputs. 

One key advantage of the property schema is that it does not rely on RDF Schema (RDFS) metadata. In practice, this allows us to represent \emph{any} file on Linked Open Data tabularly, regardless of whether metadata is available. 

Even in the simple example of Figure \ref{proptablefig}, we note that the property table is not without its challenges. It is usually \emph{sparse}, and it may end up being \emph{broad}, for RDF datasets with many properties. Furthermore, properties are \emph{named} using different conventions (for example, the prefix \emph{Has} occurs in all the properties in Figure \ref{proptablefig}) and could be \emph{opaque}, depending on the dataset publisher. We show empirically (Section \ref{experiments}) that the \emph{instantiated} pipeline (Section \ref{unsup}) can handle these difficulties. 

\subsection{Extending the formalism}\label{dnf}
The formalism in Section \ref{prelims} assumed structural homogeneity. We extend it to accommodate structural heterogeneity. 
As input, consider two datasets $R_1$ and $R_2$. If either dataset is in RDF, we assume that it is in property table form. Consider the original definition of SBPs in Definition \ref{sbp}. SBPs are associated with a \emph{single} GBP and a \emph{single} field, making them amenable only to a single schema. To account for heterogeneity, we extend SBPs by replacing the \emph{field} input with a \emph{mapping}. Denote as $A_1$ and $A_2$ the respective sets of fields of datasets $R_1$ and $R_2$. We define  \emph{simple extended} SBPs below:
\begin{definition}\label{esbp}
A \emph{simple extended specific blocking predicate} is a pair $(p_i,m)$ where $p_i$ is a general blocking predicate and $m=(\{f_1\},\{f_2\})$ is a mapping from a single field $f_1 \in A_1$ to a single field $f_2 \in A_2$. The predicate takes two tuples $t_1$ and $t_2$ as arguments and applies $p_i$ to the field values corresponding to $f_1$ and $f_2$ in the two tuples respectively.
\end{definition} 

The correspondence in Definition \ref{esbp} is denoted as \emph{simple} since it uses a mapping of the simplest cardinality (1:1). Definition \ref{sbp} can be \emph{reformulated} as a special case of Definition \ref{esbp}, with $A_1=A_2$ and $f_1=f_2$.  

The SBP semantics are not evident if the mapping cardinality is n:m, 1:n or n:1, that is, between two \emph{arbitrary} field-subsets, $F_1 \subseteq A_1$ and $F_2 \subseteq A_2$. If we \emph{interpret} the two sets as representing $|F_1||F_2|$ 1:1 mappings, we obtain a \emph{set} of simple extended SBPs $\{(p_i,(\{f_1\},\{f_2\})) | f_1 \in F_1, f_2 \in F_2\}$. 

The interpretation above is motivated by the requirement that an SBP should \emph{always} return a boolean value. We approach the problem by using the n:m mappings to construct the set of simple extended SBPs, as shown above. We then use \emph{disjunction} as a \emph{combine} operator on all elements of the constructed set to yield a single boolean result. We can then define \emph{complex extended} SBPs. 
\begin{definition}\label{cesbp}
A \emph{complex extended specific blocking predicate} is a pair $(p_i,M)$ where $p_i$ is a general blocking predicate and $M$ is a mapping from a set $F_1 \subseteq A_1$ to a set $F_2 \subseteq A_2$. The predicate takes two tuples $t_1$ and $t_2$ as arguments and applies on them $|F_1||F_2|$ simple extended SBPs $(p_i,m)$, where $m$ ranges over all 1:1 mappings in the set  $\{(\{f_1\},\{f_2\}) | f_1 \in F_1, f_2 \in F_2\}$. The predicate returns the disjunction of these $|F_1||F_2|$ values as the final output. 
\end{definition} 

\begin{ex}\label{csbpEx}
Consider the mapping between sets  \{Name\} in Dataset 2 and \{First Name, Last Name\} in Dataset 3, in Figure \ref{datasetsfig}. Let the input GBP be \emph{ContainsCommonToken}. The complex extended SBP corresponding to these inputs would be \emph{ContainsCommonToken}(\{Name\}, \{First Name\}) $\vee$ \emph{ContainsCommonToken}(\{Name\}, \{Last Name\}). This complex extended SBP would yield the same result as the simple extended SBP \emph{ContainsCommonToken}(\{Name\},\{Na-me\}) if a new field called \emph{Name} is \emph{derived} from the merging of the \emph{First Name} and \emph{Last Name} fields in Dataset 3.
\end{ex}

An operator like \emph{conjunction} is theoretically possible but may prove restrictive when learning practical schemes. The \emph{disjunction} operator makes complex extended SBPs more expressive than simple extended SBPs, but requires more computation (Section \ref{leds}). Evaluating alternate combine operators is left for future work.

Finally, (simple or complex) \emph{extended} DNF schemes can be defined in a similar vein as Definition \ref{dnfbs}, using (simple or complex) extended SBPs as atoms. One key advantage of using disjunction as the combine operator in Definition \ref{cesbp} is that, assuming simple extended SBPs as atoms for \emph{both} simple and complex extended DNF schemes, the scheme remains a positive boolean formula, by virtue of \emph{distributivity} of conjunction and disjunction. 
\subsection{Extending Existing DNF-BSLs}\label{leds}
Existing DNF-BSLs \cite{bilenkoblocking, knoblockblocking, mayankblocking, semisup} rely on similar high-level principles, which is to devise an approximation algorithm for solving the NP-hard optimization problem first formalized by Bilenko et al. \cite{bilenkoblocking}. The approximation algorithms are different in that they require different parameters and levels of supervision. These are detailed below and summarized in Table \ref{baseline}. These BSLs were originally designed only for structurally homogeneous tables, with a single field-set $A$. We describe their underlying core principles before describing extensions in keeping with the formalism in Section \ref{dnf}. 

Assume a set of GBPs G. The \emph{core} of all approximation algorithms would first construct a \emph{search space} of SBPs $H$ by forming the cross product of $G$ and $A$. The goal of the algorithm is to choose a subset $H' \subseteq H$ such that the optimization condition\footnote{We describe the condition here intuitively, and formally state it in the Appendix.} laid out by Bilenko et al. is satisfied, at least approximately \cite{bilenkoblocking}. The condition assumes that training sets of duplicates $D$ and non-duplicates $N$ are provided. Intuitively, the condition states that the \emph{disjunctive} blocking scheme formed by taking the disjunction of SBPs in $H'$ \emph{covers} (see Definition \ref{dnfbs}) at least $\epsilon|D|$ duplicates, while covering the \emph{minimum} possible non-duplicates \cite{bilenkoblocking}. Note that $\epsilon$ is a parameter common to all four systems\footnote{$\epsilon$ was designated as \emph{min\_thresh} in the original System 1 paper \cite{knoblockblocking}, and $\sigma$ in the System 3 paper \cite{semisup}.} in Table \ref{baseline}.
\begin{figure}
\centering
\epsfig{file=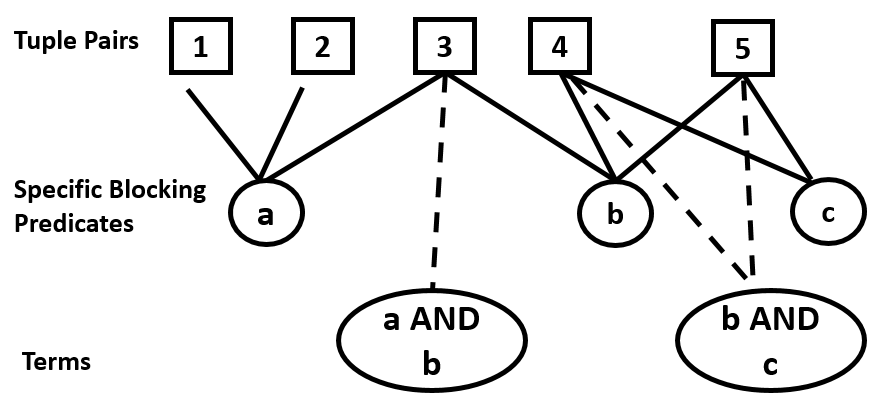, height=0.9in, width=3.0in}
\caption{An example showing how $H_c$ is formed}
\label{killfig}
\end{figure}

In order to learn a \emph{DNF} scheme (as opposed to just disjunctive), a \emph{beam search} parameter $k$ (also common to all four systems) is required. This parameter is used to \emph{supplement} the original set $H$ with \emph{terms}, to obtain a new set $H_c$. This combinatorial process is demonstrated in Figure \ref{killfig}.

$H$ originally consists of the SBPs $a$, $b$ and $c$. These SBPs cover some tuple pairs (TPs). Suppose $k=2$. A term of size 2 is formed by checking if any TP is covered by (at least) two SBPs. For example, TP-3 is covered by SBPs $a$ and $b$, and hence, also covered by the term $a \wedge b$. For $k > 2$, terms from size 2 to size $k$ are \emph{recursively} added to $H$; the final supplemented set is denoted as $H_c$. Note that for $|H|$ predicates, building $H_c$ takes $O(|H|^k)$ time \emph{per} TP. Given the exponential dependence on $k$ and diminishing returns, previous results capped $k$ at 2 \cite{mayankblocking,bilenkoblocking}. If $k=1$, $H_c=H$.    

The set $H' \subseteq H_c$ that is now chosen by the approximation scheme would potentially consist of terms and SBPs, with their disjunction yielding a k-DNF scheme\footnote{A k-DNF formula has at most k literals in each term.}.

While System 1 only requires $\epsilon$ and $k$ as its parameters, Systems 2 and 4 \emph{prune} their search spaces by removing all SBPs and terms from $H_c$ that cover more than $\eta|N|$ non-duplicates. Note that this step \emph{heuristically} improves both\footnote{Since $H_c$ now contains only a few, high-quality elements.} quality and run-time. It comes at the risk of failure, since if the search space is pruned excessively (by high $\eta$), it may become impossible to cover at least $\epsilon|D|$ duplicates. Systems 3 and 4 require less supervision but significantly more \emph{parameter tuning}, given they rely on many more parameters. 

Systems 1-3 rely on different\footnote{Surveyed briefly in the Appendix.} \emph{Set Covering} (SC) variants \cite{chvatal}. All three systems can be extended by constructing a search space of complex extended SBPs using $G$ and the mappings set $Q$, instead of $G$ and a field set $A$. The underlying SC approximations operate in this abstract search space to choose the final set $H'$. An extended DNF scheme is formed by a disjunction of (extended) elements in $H'$.   
\begin{table}
\caption{ DNF-BSL Systems}
    \begin{tabular}[t]{|p{0.2cm} |p{2.8cm} | p{2.2cm} | p{1.8cm} |}
    \hline
  {\bf ID} & {\bf System}&{\bf Parameters}&{\bf Supervision} \\ \hline
{1}&{Michelson and Knoblock \cite{knoblockblocking}} &{$\epsilon$, $k$} & {Supervised} \\ \hline
{2}&{Bilenko et al. \cite{bilenkoblocking}} &{$\epsilon$,$\eta$,$k$} & {Supervised} \\ \hline
{3}&{Cao et al. \cite{semisup}} &{$s$,$\epsilon$,$\tau$,$\alpha$,$k$} & {Semi-sup.} \\ \hline
{4}&{Kejriwal and Miranker \cite{mayankblocking}} &{{\bf Generator:} $c$,$ut$,$lt$,$d$,$nd$ {\bf Learner:} $\epsilon$,$\eta$,$k$} & {Unsup.} \\ \hline
{5}&{Algorithm \ref{alg} \emph{herein}} &{$\kappa$, $k$} & {Unsup.} \\ \hline
    \end{tabular}\label{baseline}
\end{table}

Modifying System 4  is problematic because the system is \emph{unsupervised} and runs in \emph{two} phases. The first phase (denoted as \emph{generator}) generates a noisy training set, and the second phase (denoted \emph{learner}) performs feature-selection on the noisy set to output a DNF blocking scheme. The feature-selection based learner is similar to Systems 1-3 and \emph{can} be extended. Unfortunately, the generator \emph{explicitly} assumes homogeneity, and cannot be directly extended to generate training examples for heterogeneous datasets. This implies that the DNF-BSL component of the proposed generic pipeline in Figure \ref{pipelinefig}(a) \emph{cannot} be instantiated with an existing unsupervised system.

In the event that a schema matcher (and thus, $Q$) is unavailable in Figure \ref{pipelinefig}(a), we present a \emph{fall-back option}. Specifically, we build a set $Q$ of \emph{all} 1:1 mappings, $|Q|=|A_1||A_2|$. Recall $A_i$ is the field set of dataset $i$. We denote the constructed set $H$ of SBPs as \emph{simple exhaustive}. Note that for the set of \emph{all} mappings ($\approx 2^{|A_1|}2^{|A_2|}$), the constructed set $H$ (denoted \emph{complex exhaustive}) is not computationally feasible for non-trivial cases. Even the simple exhaustive case is only a fall-back option, since a \emph{true} set of 1:1 mappings $Q$ would be much smaller\footnote{At most \emph{min($|A_1|,|A_2|$)}} than this set.

\section{An Unsupervised Instantiation}\label{unsup}

A key question to ask is whether the generic pipeline can be instantiated in an \emph{unsupervised} fashion. As we showed earlier, existing DNF-BSLs that can be extended require some form of supervision. An unsupervised heterogeneous DNF-BSL is important because, \emph{in principle}, it enables a fully unsupervised \emph{ER workflow} in both the relational and Semantic Web communities. As the surveys by Elmagarmid et al. and Ferraram et al. note, unsupervised techniques for the \emph{second} ER step do exist already \cite{recordlinkagesurvey,ERsurvey}. 
A second motivation is the observation that existing unsupervised and semi-supervised homogeneous DNF-BSLs (Systems 3-4) require considerable parameter tuning. Parameter tuning is increasingly being cited as an important algorithmic issue, in applications ranging from schema matching \cite{tuner} to generic machine learning \cite{tunerML}. \emph{Variety} in Big Data implies that algorithm design cannot discount parameter tuning. 

We propose an unsupervised instantiation with a \emph{new} DNF-BSL that requires only \emph{two} parameters. In Table \ref{baseline}, only the supervised System 1 requires two parameters. The schematic of the unsupervised instantiation (of the generic pipeline in Figure \ref{pipelinefig}(a)) is shown in Figure \ref{pipelinefig}(b). 
We use the \emph{existing} schema matcher, \emph{Dumas}, in the instantiated pipeline \cite{dumas}. Dumas outputs 1:1 field mappings by first using a \emph{duplicates generator} to locate  tuple pairs with high \emph{cosine similarity}. In the second step, Dumas uses \emph{Soft-TFIDF} to build a \emph{similarity matrix} from each generated duplicate. If $n$ duplicates are input to the second step, $n$ similarity matrices are built and then averaged into a single similarity matrix. The \emph{assignment problem} is then solved by invoking the \emph{Hungarian Algorithm} on this matrix \cite{hungarian}. This results in exactly $min(|A_1|,|A_2|)$ 1:1 field mappings (the set $Q$) being output.

In addition to using $Q$, we recycle the noisy duplicates of Dumas and pipe them into Algorithm \ref{alg}. Note that Dumas does not generate non-duplicates. We address this issue in a novel way, by \emph{permuting} the generated duplicates set $D$. Suppose that $D$ contains $n$ tuple pairs $\{(r_1,s_1),\ldots(r_n,s_n)\}$, with each $r,s$ respectively from datasets $R_1,R_2$. By randomly permuting the pairs in $D$, we \emph{heuristically} obtain non-duplicate pairs of the form $(r_i,s_j)$, $i \neq j$. Note that (at most) $n!$ distinct permutations are possible. For balanced supervision, we set $|N|=|D|$, with $N$ the permutation-generated set.  

Empirically, the permutation is expected to yield a precise $N$ because of observed \emph{duplicates sparsity} in ER datasets \cite{christensurvey,mayankblocking}. This sparsity is also a key tenet underlying the blocking procedure itself. If the datasets were dense in duplicates, blocking would not yield any savings. 

\begin{algorithm}
\caption{Learn Extended k-DNF Blocking Scheme}
\begin{algorithmic}\label{alg}
\STATE \textbf{Input :} Set $D$ of duplicate tuple pairs, Set $Q$ of mappings
\STATE \textbf{Parameters :} Beam search parameter $k$, SC-threshold $\kappa$
\STATE \textbf{Output :} Extended DNF Blocking Scheme $\mathcal{B}$
\STATE \textbf{Method :}
\emph{//Step 0: Construct sets $N$ and $H$}
\STATE{\emph{Permute} pairs in $D$ to obtain $N$, $|N|=|D|$}
\STATE{Construct set $H$ of simple extended SBPs using set $G$ of GBPs and $Q$}
\STATE{Supplement set $H$ to get set $H_c$ using $k$}\\
\emph{//Step 1: Build Multimaps $M'_D$ and $M'_N$} \\
\STATE{Construct $M_D=<X,H_X>$, X is a tuple pair in $D$, $H_X \subseteq H_c$ contains the elements in $H_c$ \emph{covering} X}
\STATE{Repeat previous step to build $M_N$ for tuple pairs in $N$}\\
\STATE{ \emph{Reverse} $M_D$ and $M_N$ to respectively get $M'_D$ and $M'_N$}\\
\emph{//Step 2: Run approximation algorithm}
\FORALL{$X \in keyset(M'_D)$}
\STATE{\emph{Score} $X$ by using formula $|M'_D(X)|/|D|-|M'_N(X)|/|N|$}
\STATE{Remove $X$ if $score(X) < \kappa$}
\ENDFOR
\STATE{Perform W-SC on keys in $M'_D$ using Chvatal's heuristic, weights are \emph{negative} scores}\\
\emph{//Step 3: Construct and output DNF blocking scheme}
\STATE{$\mathcal{B}:=$ Disjunction of chosen keys} 
\STATE{Output $\mathcal{B}$}
\end{algorithmic}
\end{algorithm}  

Algorithm \ref{alg} shows the pseudocode of the extended DNF BSL. Inputs to the algorithm are the piped Dumas outputs, $D$ and $Q$.  To learn a blocking scheme from these inputs, two parameters $k$ and $\kappa$ need to be specified. Similar to (extended) Systems 1-3 in Table \ref{baseline}, $G$, $Q$ and $k$ are used to construct the search space, $H_c$. Note that $G$ is considered the algorithm's \emph{feature space}, and is not a dataset-dependent input (or parameter). Designing an \emph{expressive} $G$ has computational and qualitative effects, as we empirically demonstrate. We describe the GBPs included in $G$ in Section \ref{experiments}.   
 
Step 0 in Algorithm \ref{alg} is the permutation step just described, to generate the non-duplicates set $N$. $G$ and $Q$ are then used to construct the set $H$ of simple extended\footnote{Since Dumas only outputs 1:1 mappings} SBPs (Definition \ref{esbp}), with $|H|=|G||Q|$. $H$ is supplemented (using parameter $k$) to yield $H_c$, as earlier described in Section \ref{leds}. 

\begin{figure}
\centering
\epsfig{file=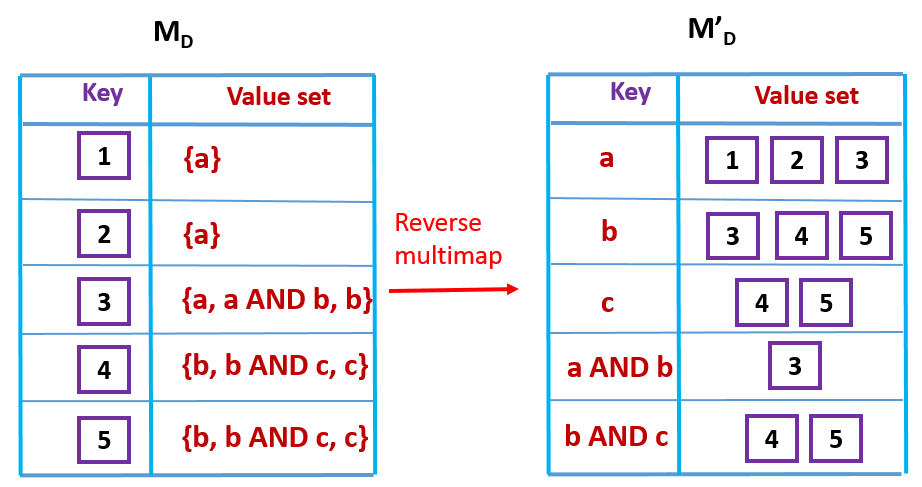, height=1.3in, width=3.3in}
\caption{Step 1 of Algorithm \ref{alg}, assuming the information in Figure \ref{killfig}}
\label{mapfig}
\end{figure} 
Step 1 constructs \emph{multimaps}\footnote{Multimap keys reference multiple values (or a \emph{value set}).} on which Set Covering (SC) is eventually run. As a first logical step, multimaps $M_D$  and $M_N$ are constructed. Each tuple pair (TP) in $D$ is a \emph{key} in $M_D$, with the SBPs and terms in $H_c$ covering that TP comprising the \emph{value set}. $M_D$ is then \emph{reversed} to yield $M'_D$. $M'_N$ is built analogously. Figure \ref{mapfig} demonstrates the procedure, assuming $D$ contains TPs 1-5, covered as shown in Figure \ref{killfig}. The time complexity of building (both) $M'_D$ and $M'_N$ is $O(|H|^k(|D|+|N|))$.  

In Step 2, each key is first scored by calculating the difference between the \emph{fractions} of covered duplicates and non-duplicates. A threshold parameter, $\kappa$, is used to remove the SBPs and terms that have low scores. Intuitively, $\kappa$ tries to \emph{balance} the conflicting needs of previously described parameters, $\epsilon$ and $\eta$, and reduce tuning effort. The range of $\kappa$ is [-1,1]. An advantage of the parameter is that it has an intuitive interpretation. A value close to $1.0$ would indicate that the user is confident about low noise-levels in inputs $D$ and $Q$, since high $\kappa$ implies the existence of elements in $H_c$ that cover many positives and few negatives. Since many keys in $M'_D$ are removed by high $\kappa$, this also leads to computational savings. However, setting $\kappa$ too high (perhaps because of misguided user confidence) could potentially lead to excessive purging of $M'_D$, and subsequent algorithm failure. Experimentally, we show that $\kappa$ is easily tunable and even high values of $\kappa$ are robust to noisy inputs. 

\emph{Weighted Set Covering} (W-SC) is then performed using \emph{Chvatal's algorithm}\footnote{We include Chvatal's algorithm in the SC Appendix survey.} \cite{chvatal}, with each key in $M'_D$ acting as a \emph{set} and the tuple pairs covered by all keys as elements of the \emph{universe set} $\mathcal{U}$. For example, assuming all SBPs and terms in the keyset of $M'_D$ in Figure \ref{mapfig} have scores above $\kappa$, $\mathcal{U}= \{1,2,3,4,5\}$. Note that \emph{only} $M'_D$ is pruned (using $\kappa$) and also, W-SC is performed only on $M'_D$. $M'_N$ only aids in the score calculation (and subsequent pruning process) and may be safely purged from memory before W-SC commences.

W-SC needs to find a subset of the $M'_D$ \emph{keyset} that covers \emph{all} of $\mathcal{U}$ and with \emph{minimum} total weight. For this reason, the weight of each set is the \emph{negative} of its calculated score. Given that sets chosen by W-SC actually represent SBPs or terms, their disjunction is the  k-DNF blocking scheme.   

Under plausible\footnote{$P \subset NP$} complexity assumptions, Chvatal's algorithm is essentially the \emph{best-known} polynomial-time approximation for W-SC \cite{setcoverguarantee}. For example, Bilenko et al. used Peleg's approximation to \emph{Red-Blue SC} \cite{approx, redblue}, which is known to have worse bounds \cite{redblue}. The proposed DNF-BSL has the strongest theoretical approximation guarantee of all systems in Table \ref{baseline}.   

\section{Experiments}\label{experiments}

\subsection{Metrics}
The quality and efficiency of blocking is evaluated by the special metrics, \emph{Reduction Ratio} (RR), \emph{Pairs Completeness} (PC) and \emph{Pairs Quality} (PQ) \cite{christensurvey}. Traditional metrics like precision and recall do not apply to blocking since it is a \emph{pre-processing} step, and its output is not the final ER output. To define RR, which intuitively measures \emph{efficiency}, consider the full set of pairs $\Omega$ that would be generated in the \emph{absence} of blocking. Specifically, $\Omega$ is the set $\{(r,s)|r \in R_1, s \in R_2  \}$. RR is then given by the formula, $1-|\Gamma|/|\Omega|$. Given the sparsity of duplicates in real-world datasets \cite{christensurvey}, RR should ideally be close to, but not precisely, $1.0$ (unless $\Gamma=\phi$).

Denote the set of all duplicate pairs, or the \emph{ground-truth}, as $\Omega_m$. Similarly, consider the set of duplicates \emph{included} in the candidate set $\Gamma_m = \Gamma \cap \Omega_m$. The metric PC quantifies the \emph{effectiveness} of the blocking scheme by measuring \emph{coverage} of $\Gamma_m$ with respect to $\Omega_m$. Specifically, it is given by the formula, $|\Gamma_m|/|\Omega_m|$. Low PC implies that recall on \emph{overall} ER will be low, since many duplicates did not share a block to begin with. 

PC and RR together express an efficiency-effectiveness tradeoff. The metric PQ is sometimes used to measure how \emph{dense} the blocks are in duplicates, and is given by $|\Gamma_m|/|\Gamma|$. PQ has not been reported in recent BSL literature \cite{bilenkoblocking},\cite{mayankblocking}. One reason is that PQ can be equivalently expressed as $c.PC/(1-RR)$, where $c$ is $|\Omega_m|/|\Omega|$. When \emph{comparing} two BSLs, PQ can therefore be expressed wholly in terms of PC and RR. We do not consider PQ further in this paper. 

\subsection{Datasets}
\begin{table*}[!t]
\caption{Details of dataset pairs. The notation, where applicable, is (first dataset) /$\times$ (second dataset) }
    \begin{tabular}[t]{|p{0.5cm} |p{3.9cm} | p{0.8cm} | p{4.5cm} |p{2.5cm}|p{2.5cm}|}
\hline
    {\bf ID}& {\bf Dataset Pairs}&{\bf Fields}&{\bf Total Entity Pairs} & {\bf Duplicate Pairs}& {\bf Data Model}\\ \hline
{1}&{Restaurant 1 /Restaurant 2} & {8/8} & {339 $\times$ 2256 = 764,784} & {100} & {RDF/RDF} \\ \hline
{2}&{Persons 1 /Persons 2} & {15/14} & {2000 $\times$ 1000 = 2 million} & {500} & {RDF/RDF} \\ \hline
{3}&{IBM/vgchartz} & {12/11} & {1904 $\times$ 20,000 $\approx$ 38 million} & {3933} & {Tabular/Tabular} \\ \hline
{4}&{Libraries 1 /Libraries 2} & {5/10} & {17,636 $\times$ 26,583 $\approx$ 469 million} & {16,789} & {Tabular/Tabular} \\ \hline
{5}&{IBM/DBpedia} & {12/4} & {1904 $\times$ 16,755 $\approx$ 32 million} & {748} & {Tabular/RDF} \\ \hline

{6}&{vgchartz/DBpedia} & {11/4} & {20,000 $\times$ 16,755 $\approx$ 335 million} & {10,000} & {Tabular/RDF} \\ \hline

    \end{tabular}
\label{datasets}
\end{table*}

The heterogeneous test suite of six dataset \emph{pairs} (and nine individual datasets) is summarized in Table \ref{datasets}. The suite spans over four domains and the three kinds of heterogeneity discussed in the paper. All datasets are from real-world sources. We did not curate these files in any way, except for serializing RDF datasets as property tables (instead of triples-sets). The serializing was found to be near-instantaneous (< 1 second) in all cases, with negligible run-time compared to the rest of the pipeline. 

\emph{Dataset Pairs} (DPs) 1 and 2 are the RDF benchmarks in the 2010 \emph{instance-matching} track\footnote{\url{http://oaei.ontologymatching.org/2013/#instance}} of OAEI\footnote{Ontology Alignment Evaluation Initiative}, an annual Semantic Web initiative. Note that an earlier \emph{tabular} version of DP 1 is also popular in \emph{homogeneous} ER literature \cite{christensurvey}. 

DPs 3,5 and 6  describe \emph{video game} information. DP 6 has already been used as a test case in a previous schema matching work \cite{tian}. \emph{vgchartz} is a tabular dataset taken from a reputable charting website\footnote{\url{vgchartz.com}}. \emph{DBpedia} contains 48,132 \emph{triples} extracted from DBpedia\footnote{\url{dbpedia.org}}, and has four (three\footnote{\emph{genre}, \emph{platform} and \emph{manufacturer}.} properties and \emph{subject}) fields and 16,755 tuples in property table form. Finally, \emph{IBM} contains user-contributed data extracted from the \emph{Many Eyes} page\footnote{\url{www-958.ibm.com/software/data/cognos/manyeyes/datasets}}, maintained by IBM Research. 

DP 4 describes US \emph{libraries}. \emph{Libraries 1} was from a \emph{Point of Interest} website\footnote{\url{http://www.poi-factory.com/poifiles}}, and \emph{Libraries 2} was taken from a US government listing of libraries. 

 DPs 2 and 4 contain n:m ground-truth schema mappings, while the others only contain 1:1 ground-truth mappings.

\subsection{Baselines}

Table \ref{baseline} listed existing DNF-BSLs, with Section \ref{leds} describing how the extended versions fit into the pipeline. We extend two state-of-the-art systems as baselines: 

{\bf Supervised Baseline:} System 2 was chosen as the supervised baseline \cite{bilenkoblocking}, and favored over System 1 \cite{knoblockblocking} because of better reported empirical performance on a common benchmark employed by both efforts. Tuning the parameter $\eta$ leads to better blocking schemes, making System 2 a state-of-the-art supervised baseline.  

{\bf Semi-supervised Baseline: }We \emph{adapt} System 4 as a \emph{semi-supervised} baseline by feeding it the same noisy duplicates generated by Dumas as fed to the proposed learner, as well as a \emph{manually} labeled negative training set. The learner of System 4 uses \emph{feature selection} and was shown to be empirically competitive with \emph{supervised} baselines \cite{mayankblocking}. In contrast, System 3 did not evaluate its results against System 2. Also, the learner of System 4 requires three parameters, versus the five of System 3. Finally, the three System 4 parameters are \emph{comparable} to the corresponding System 2 parameters, and evaluations in the original work showed that the learner is robust to minor parameter variations \cite{mayankblocking}. For these reasons, the feature-selection learner of System 4 was extended to serve as a semi-supervised baseline.

\subsection{Methodology}
In this section, we describe experimental methodology. Experimental \emph{results} will be presented in Section \ref{results}.
\subsubsection{Dumas}
Given that the test suite is larger and more heterogeneous in this paper than in the original Dumas work \cite{dumas}, we perform two \emph{preliminary} experiments to evaluate Dumas. Dumas was briefly described in Section \ref{unsup}.

{\bf Preliminary Experiment 1:} We evaluate the performance of Dumas's \emph{duplicates generator}. The generator uses TF-IDF to retrieve the highest scoring duplicates \emph{in order}. Suppose we retrieve $t$ duplicates (as an ordered list). Denote, for any $k \leq t$, the number of true positives in sub-list $[1\ldots k]$ as $d(k)$. Define \emph{Precision@k} as $d(k)/k$ and \emph{Recall@k} as $d(k)/|\Omega_m|$, where $\Omega_m$ is the ground-truth set of all duplicates.  
 We plot \emph{Precision@k} against \emph{Recall@k} for all DPs to demonstrate the precision-recall tradeoff. To obtain a full set of data points, we set $t$ at 50,000 for all experiments, and calculate \emph{Precision@k} and \emph{Recall@k} for $k \in \{1 \ldots t\}$. 

{\bf Preliminary Experiment 2:} Although $t$ was set to a high value in the previous experiment, Dumas requires only a few top pairs (typically 10-50) for the schema matching step \cite{dumas}. To compute the similarity matrix from $t$ pairs, Dumas uses \emph{Soft} TF-IDF, which requires an \emph{optional} threshold $\theta$. 
For a given DP, denote $Q'$ as the ground-truth set of schema mappings, $Q$ as the set of Dumas mappings, and $Q_m \subseteq Q$ as the set of \emph{correct} Dumas mappings. Define \emph{precision} (on this task) as $|Q_m|/|Q|$ and \emph{recall} as $|Q_m|/|Q'|$. In this experiment, we set $t$ and $\theta$ to \emph{default} values of $50$ and $0.5$ respectively and report the results. We also vary these parameters and describe performance differences. 

\subsubsection{DNF BSL Parameter Tuning}
We describe parameter tuning methodology. Note that the beam search parameter $k$ (see Table \ref{baseline}) is not technically tuned, but \emph{set}, incumbent on the experiment (Section \ref{dnfexperiments}).

{\bf Baseline Parameters:} Both baseline parameters are tuned in similar ways. We do an exhaustive parameter \emph{sweep} of $\epsilon$ and $\eta$, with values in the range of $0.90-0.95$ and $0.0-0.05$ (respectively) typically maximizing baseline performance. Low $\eta$ maximizes RR, while high $\epsilon$ maximizes PC. Although extreme values, with $\eta=0$ and $\epsilon=1$, are \emph{expected} to maximize performance, they led to failure\footnote{$\epsilon|D|$ duplicates could not be covered (see Section \ref{leds}).} on all test cases. This demonstrates the necessity of proper parameter tuning. Fewer parameters imply less tuning, and faster system deployment. 

{\bf SC-threshold $\kappa$:} The single parameter $\kappa$ of the proposed method was tuned on the smallest test case (DP 1) and found to work best at $0.9$. To test the \emph{robustness} of $\kappa$ to different domains, we fixed it at $0.9$ for all experiments. 

{\bf $|D|$ and $|N|$:} We emulate the methodology of Bilenko et al. in setting the numbers of positive and negative samples input to the system. Bilenko et al. input 50\% of true positives and an \emph{equal} number of negatives to train their system \cite{bilenkoblocking}. Let this number be $n$ ($=|D|$,$|N|$) for a given test suite. For example, $n=50$ for DP 1. For fairness, we use these numbers for the semi-supervised baseline and proposed system also. We retrieve the top $n$ pairs from the Dumas generator and input the pairs to both systems as $D$. Additionally, we provide $n$ labeled non-duplicates (as $N$) to the semi-supervised baseline. In subsequent experiments, the dependence of the proposed system on $n$ is tested.

\subsubsection{Extended DNF BSL}\label{dnfexperiments}
We describe the evaluation of the extended DNF BSLs.

{\bf Set $G$ of GBPs:} Bilenko et al. first proposed a set G that has since been adopted in future works \cite{bilenkoblocking, semisup, mayankblocking}. This original set contained generic token-based functions, numerical  functions and character-based functions \cite{bilenkoblocking}. No recent work attempted to supplement $G$ with more expressive GBPs. In particular, \emph{phonetic} GBPs such as \emph{Soundex}, \emph{Metaphone} and \emph{NYSIIS} were not added to $G$, despite proven performance benefits \cite{datamatching}. We make an empirical contribution by supplementing $G$ with all nine phonetic features implemented in an open-source package\footnote{org.apache.commons.codec.language}. For fairness, the same $G$ is always input to all learners in the experiments below. Results on Experiment 2 will indirectly demonstrate the benefits of supplementing $G$. A more detailed description of the original (and supplemented) $G$ is provided in the Appendix.

{\bf Experiment 1:} To evaluate the proposed \emph{learner} against the baselines, we input the same $Q$ (found in Preliminary Experiment 2) to \emph{all} three systems. The systems learn the DNF scheme by choosing a subset $H' \subseteq H_c $ of SBPs and supplemented terms, with $H_c$ constructed from $G$, $Q$ and $k$ (Section \ref{leds}). We learn only \emph{disjunctive} schemes by setting $k$ to 1 in this experiment. 

{\bf Experiment 2:} We repeat Experiment 1 but with $k=2$. We mentioned in Section \ref{leds} that complexity is \emph{exponential} in $k$ for all systems in Table \ref{baseline}. Because of this exponential dependence, it was not possible to run the experiment for $k=2$ on all DPs. We note which DPs presented problems, and why. Note that, if $G$, $Q$ and the training sets are fixed, increasing $k$ seems to be the only feasible way of improving blocking quality. However, $G$ is more expressive in this paper. Intuitively, we expect the difference across Experiments 1 and 2 to be narrower than in previous work. 
 
{\bf Experiment 3:} In a third set of experiments, we evaluate how blocking performance varies with $Q$. To the baseline methods, we input the set of (possibly n:m) ground-truth mappings $Q'$ while the Dumas output $Q$ is retained for the proposed learner. The goal is to evaluate if extended DNF-BSLs are \emph{sensitive}, or if noisy 1:1 matchers like Dumas suffice for the end goal.

{\bf Experiment 4:} We report on run-times and show performance variations of the proposed system with the number of provided duplicates, $n$. In industrial settings with an unknown ground-truth, $n$ would have to be estimated. An important question is if we can rely on getting good results with \emph{constant} $n$, despite DP heterogeneity.  

{\bf Statistical Significance:} We conduct experiments in ten runs, and (where relevant) report statistical significance levels using the paired sample Student's t-distribution. On blocking metrics, we report if results are not significant (NS), weakly significant (WS), significant (SS) or highly significant (HS), based on whether the p-value falls within brackets [1.0, 0.1), (0.05,0.1], (0.01, 0.05] and [0.0, 0.01] respectively.  As for the choice of samples, we always \emph{individually} paired PC and RR of the proposed system against the baseline that achieved a \emph{better} average on the metric.

{\bf Implementation: }All programs were implemented in Java on a 32-bit Ubuntu virtual machine with 3385 MB of RAM and a 2.40 GHz Intel 4700MQ i7 processor.

\section{Results and Discussion}\label{results}
\subsection{Dumas}
\begin{figure}
\centering
\epsfig{file=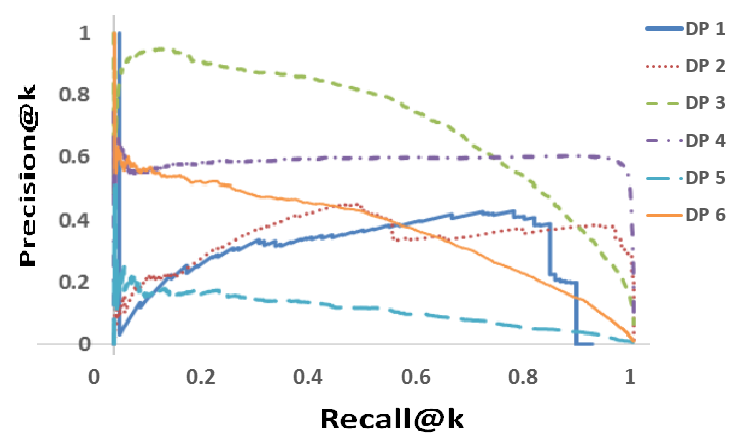, height=1.45in, width=2.8in}
\caption{Dumas duplicates-generation results, for $k \in \{1,2 \ldots 50000\}$}
\label{dumas1}
\end{figure}
{\bf Preliminary Experiment 1:} Figure \ref{dumas1} shows the results of Dumas's duplicates-generator. Except for DP 3, precision on all cases seems inadequate even at low recall levels. Although recall of 100\% is eventually attained on most DPs, the price is (near) 0\% precision. Closer inspection of the results showed that many false positives got ranked at the top. We discuss the implications of these noisy results shortly.
\begin{table}
\caption{Best Results of Dumas Schema-Matcher}
    \begin{tabular}[t]{|p{2.2cm} |p{2.4cm} | p{1.8cm} |}
    \hline
  {\bf Dataset Pair} & {\bf  Precision}&{\bf Recall} \\ \hline
{1}&{100\%} &{87.5\%}  \\ \hline
{2} & {92.86\%} & {86.67\%} \\ \hline
{3}&{91\%} &{100\%}  \\ \hline
{4}&{75\%} &{33.33\%}  \\ \hline
{5}&{100\%} &{100\%}  \\ \hline
{6}&{100\%} &{100\%}  \\ \hline
    \end{tabular}
\label{systems}
\end{table}

{\bf Preliminary Experiment 2:} Table \ref{systems} shows the schema mapping results retrieved from Dumas with parameters set at $t=50$ and $\theta=0.5$. These default values were found to maximize performance in \emph{all} cases, in agreement with similar values in the original work \cite{dumas}. We also varied $t$ from 10 to 10,000 and $\theta$ from 0 to 0.9. Performance slightly declined (by at most 5\%) for some DPs when $t < 50$, but remained otherwise constant across parameter sweeps. This confirms that Dumas is quite robust to $t$ and $\theta$. 
One \emph{disadvantage} of Dumas is that it is a 1:1 matcher. This explains the lower recall numbers on DPs 2 and 4, which contain n:m mappings. In Experiment 3, we test if this is problematic by providing the ground-truth set $Q'$ to baselines, and comparing results. 

{\bf Discussion:} An important point to note from the (mostly) good results in Table \ref{systems} is that generator accuracy is not always \emph{predictive} of schema mapping accuracy. This robustness to noise is always an important criteria in pipelines. If noise \emph{accumulates} at each step, the final results will be qualitatively weak, possibly meaningless. Since Dumas's matching component is able to compensate for generator noise, it is a good empirical candidate for unsupervised 1:1 matching on typical ER cases. Preliminary Experiment 1 also shows that on real-world ER problems, a simple measure like TF-IDF is not appropriate (by itself) for solving ER. The generator noise provides an interesting test for the extended DNF-BSLs. Since both the mappings-set $Q$ and top $n$ generated duplicates (the set $D$) output by Dumas are piped to the learner, there are \emph{two} potential sources of noise.    

Finally, given that the proposed system (in subsequent experiments) \emph{permutes} the Dumas-output $D$ (Algorithm \ref{alg}) to generate $N$, we tested the accuracy of the permutation procedure in a follow-up experiment.  With $|D|$ ranging from 50 to 10,000 in increments of 50, we generated $N$ of equal size and calculated \emph{non-duplicates accuracy}\footnote{$1-|N \cap \Omega_m|/|N|$} of $N$ over ten random trials per value. In all cases, accuracy was 100\%, showing that the permutation heuristic is actually quite strong.
\subsection{Extended DNF BSL}
\begin{table*}[!t]
{
\caption{Comparative Results of Extended DNF BSLs. Bold values are (at least) weakly significant, with significance levels (WS,SS or HS) in paranthesis}
  
 \begin{tabular}[t]{ |p{0.7cm} | p{3cm} | p{2.2cm} | p{1.2cm} | p{1.2cm} | p{2.1cm} | p{2.1cm} | p{2.1cm} | }
    \hline
   \multicolumn{2}{|c|}{\multirow{2}{*}{Dataset Pair (DP)}} & \multicolumn{2}{ |c| }{Proposed Method} & \multicolumn{2}{ |c| }{Semi-Supervised Baseline} & \multicolumn{2}{ |c| }{Supervised Baseline} \\ \cline{3-8} 
& &PC & RR &PC & RR  &PC & RR \\ \hline
\multirow{2}{*}{DP 1} & Average & 100\%  & 99.68\%  &  100\% & 99.68\%  &  100\% & 98.59\% \\ \cline{3-8}
 & Standard Deviation& 0\% & 0\% & 0\% & 0\%  &  0\% & 3.41\% \\ \hline
\multirow{2}{*}{DP 2} & Average & 95\%  & 86.11\%  &  95\% & 99.23\%  &  {\bf 99.6\% (HS)} & {\bf 99.96\% (HS)}  \\ \cline{3-8}
 & Standard Deviation& 0\% & 0\% & 0\% & 0\%  &  0\% & 0\% \\ \hline
\multirow{2}{*}{DP 3} & Average &  100\%  &95.47\%  &  100\% & 95.44\%    & 99.29\% & {\bf 99.99\% (HS)}\\ \cline{3-8}
 & Standard Deviation & 0\% & 0\% & 0\% &0.01\%  &  0\% & 0\% \\ \hline
\multirow{2}{*}{DP 4} & Average & {\bf 98.95\% (HS)} & 99.68\%  &  98.43\% & {\bf 99.98\% (HS)}    &  98.19\% & {\bf 99.98\% (HS)}\\ \cline{3-8}
 & Standard Deviation& 0.1\% & 0.007\% & 0\% & 0\%  &  0.27\% & 0.01\% \\ \hline
\multirow{2}{*}{DP 5} & Average & 100\%  & 92.28\% & 100\% & 94.58\%  &  99.46\% & {\bf 97.75\% (HS)}  \\ \cline{3-8}
 & Standard Deviation & 0\% & 0\%  &  0\%  & 1.01\%  &0.8\% & 2.36\% \\ \hline
 \multirow{2}{*}{DP 6} & Average & 99.91\%  & 99.69\%  &   99.97\% & 99.71\%    &  91.09\% & {\bf 99.93\% (HS)}\\ \cline{3-8}
 &Standard Deviation & 0.08\% & 0.019\% & 0.07\% & 0.02\%  &  0.03\% & 0.004\% \\ \hline
    \end{tabular}\label{block}}
\end{table*}
{\bf Experiment 1:} Table \ref{block} shows BSL results on all six DPs. The high overall performance explains the recent popularity of DNF-BSLs. Using the extended DNF hypothesis space for blocking schemes allows the learner to compensate for the two sources of noise earlier discussed.
Overall, considering statistically significant results, the supervised method typically achieves better RR, but PC is (mostly) equally high for all methods, with the proposed method performing the best on DP 4 (the largest DP) and the supervised baseline on DP 2, with high significance. We believe the former result was obtained because the proposed method has the strongest approximation bounds out of all three systems, and that this effect would be most apparent on large DPs.  Importantly, low standard deviation (often 0) is frequently observed for all methods. The DNF-BSLs prove to be quite \emph{deterministic}, which can be important when replicating results in both research and industrial settings.     

{\bf Experiment 2:} Next, we evaluated if $k=2$ enhances BSL performance and justifies the exponentially increased cost. With $k=2$, only DPs 1 and 5 were found computationally feasible. On the other DPs, the program either ran out of RAM (DPs 4,6), or did not terminate after a long\footnote{Within a factor of 20 of the average time taken by the system for the $k=1$ experiment (for that DP).} time (DPs 2,3). The former was observed because of high $n$ and the latter because of the large number of \emph{fields} (see Table \ref{datasets}). 
Setting $k$ beyond 2 was computationally infeasible even for DPs 1 and 5. Furthermore, results on DPs 1 and 5 showed \emph{no} statistical difference compared to Experiment 1, even though run-times went up by an approximate factor of 16 (for both DPs). 

{\bf Experiment 3:} We provided the ground-truth set $Q'$ to baseline methods (and with $k$ again set to 1), while retaining $Q$ for the proposed method. Again, we did not observe any statistically significant difference in PC or RR for either baseline method. We believe this is because the cases for which $Q'$ would most likely have proved useful (DPs 2 and 4, which contain n:m mappings that Dumas cannot output) already perform well with the Dumas-output $Q$.
\begin{figure*}[t]
\centering
\epsfig{file=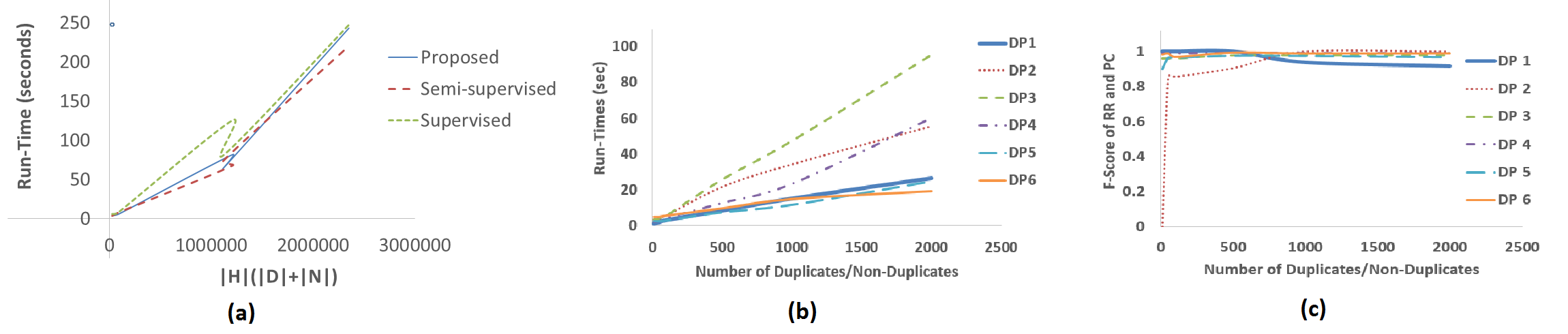, height=1.6in, width=6.5in}
\caption{Experiment 4 results. (a) plots run-time trends of all three systems against the theoretical formula, while (b) and (c) respectively plot run-times and f-scores of the proposed system against sample sizes.}
\label{combinedfig}
\end{figure*}

{\bf Experiment 4:} Theoretically, run-time of Algorithm \ref{alg} was shown to be $O(|H|^k(|D|+|N|))$; the run-times of other systems in Table \ref{baseline} are similar. Empirically, this has never been demonstrated. For $k=1$, we plot the run-time data collected from Experiment 1 runs on DPs 1-6 (Figure \ref{combinedfig}(a)). The trend is fairly linear, with the supervised system slower for smaller inputs, but not larger inputs. The dependence on $|Q|$ shows why a schema matcher is necessary, since in its absence, the \emph{simple exhaustive} set  is input (Section \ref{leds}).

As further validation of the theoretical run-time, Figure \ref{combinedfig}(b) shows the linear dependence of the proposed system on $|D|$. Again, the trend is linear, but the slope\footnote{The slope is the hidden constant in the asympotic notation.} depends on the \emph{schema heterogeneities} of the individual DPs. For example, DPs 2 and 3, the largest datasets in terms of \emph{fields} (Table \ref{datasets}), do not scale as well as the others.

Figure \ref{combinedfig}(c) shows an important \emph{robustness} result. We plot the PC-RR \emph{f-score}\footnote{$\frac{2.RR.PC}{RR+PC}$} of the proposed system against $|D|$. The first observation is that, even for small $|D|$, performance is already high except on DP 2, which shows a steep increase when $|D| \approx 100$.  On all cases, maximum performance is achieved at about $|D| \approx 700$ and the f-score curves flatten subsequently. This is qualitatively similar to the robustness of Dumas to small numbers of positive samples.

Figure \ref{combinedfig}(c) also shows that the proposed method is  robust to \emph{overestimates} of $|D|$. DP 1, for example, only has  100 true positives, but it continues to perform well at much higher $|D|$  (albeit with a slight dip at $|D| \approx 700$) . 

{\bf Discussion:} We earlier stated, when discussing the preliminary experimental results, that an extended DNF-BSL can only integrate well into the pipeline if it is robust to noise from previous steps. Previous research has noted the overall robustness of DNF-BSLs. This led to the recent emergence of a homogeneous unsupervised system \cite{mayankblocking}, adapted here as a semi-supervised baseline. Experiment 1 results showed that this robustness also carries over to \emph{extended} DNF-BSLs. High overall performance shows that the pipeline can accommodate heterogeneity, a key goal of this paper. 

Experiment 2 results demonstrate the advantage of having an expressive $G$, which is evidently more viable than increasing $k$. On DPs 1 and 5 (that the systems succeeded on), no statistically significant differences were observed, despite run-time increasing by a factor of 16. We note that the largest (homogeneous) test cases on which $k=2$ schemes were previously evaluated were only about the order of DP 1 (in size). Even with less expressive $G$, only a few percentage point performance differences were observed (in PC and RR), with statistical significance \emph{not} reported \cite{mayankblocking, bilenkoblocking}. 

To confirm the role of $G$, we performed a follow-up experiment where we used the originally proposed $G$ \cite{bilenkoblocking} on DPs 1 and 5, with both $k=1$ and $k=2$. We observed lower performance with $k=1$ compared to Table \ref{block} results, while $k=2$ results were only at par with those results. Run-times with less expressive $G$ were obviously lower (for corresponding $k$); however, $k=2$ run-times were higher (with \emph{less} expressive $G$) than $k=1$ run-times with \emph{more} expressive $G$. All differences just described were statistically significant (at the 95\% level). 
This validates previous research findings, while also confirming our stated hypothesis about $G$.

Experiment 3 results showed that a sophisticated schema matcher is not always necessary for the purpose of learning a blocking scheme. However, the importance of good schema matching goes beyond blocking and even ER. Schema matching is an important step in overall data integration \cite{schemamatching}. On noisier datasets, a good n:m schema matcher could make all the difference in pipeline performance, but we leave for future work to evaluate such a case.

The similar run-time trends shown by the various systems in Figure \ref{combinedfig}(a) also explain why, in Experiment 2, all systems simultaneously succeeded or failed on a given DP. Even if we replace our DNF-BSL with an extended version from the literature, the exponential dependence on $k$ remains. Figures \ref{combinedfig}(a) and (b) also validate theoretical run-time calculations empirically. Previous research on DNF-BSLs did not theoretically analyze (or empirically report) algorithmic run-times and scalability explicitly \cite{mayankblocking, bilenkoblocking, semisup, knoblockblocking}.  

Figure \ref{combinedfig}(c) demonstrates the encouraging qualitative result that only a few (noisy) samples are typically enough for adequate performance. Given enterprise quality requirements, as well as expense of domain expertise, high performance for low $n$ and minimum parameter tuning is a practical necessity, for industrial deployment. Recall that we retained $\kappa$ at 0.9 for \emph{all} experiments (after tuning on DP 1), while for the baselines, we had to conduct parameter sweeps for each separate experiment. Combined with results in both Table \ref{block} and Figure \ref{combinedfig}(c), this shows that the system can be a potential use-case in industry. Combined with previous unsupervised results for the \emph{second} ER step \cite{ERsurvey, recordlinkagesurvey}, such a use-case would apply both to relational and Semantic Web data as a \emph{fully unsupervised ER workflow}, which has thus far remained elusive. 
\section{Conclusion and Future Work}\label{conclusion}
In this paper, we presented a generic pipeline for learning DNF blocking schemes on heterogeneous dataset pairs. We proposed an unsupervised instantiation of the pipeline that relies on an existing instance-based schema matcher and learns blocking schemes using only two parameters.
We also showed a novel way of reconciling RDF-tabular heterogeneity by using the \emph{logical} property table representation for building and populating a  dynamic property schema for RDF datasets. Finally, we evaluated all techniques on six test cases exhibiting three separate kinds of heterogeneity.

Future research will address further exploration of the property table representation for tabularly mining RDF data. Additionally, refining $G$ (the set of GBPs) further is a promising, proven method of scalably improving BSL performance. We will also implement a fully unsupervised ER workflow that the proposed unsupervised DNF-BSL \emph{enables}, and evaluate it in a similar fashion. 
\bibliographystyle{abbrv}
\bibliography{sigproc-sp}  

\section*{APPENDIX}
\subsection*{Property Table Algorithms}
We describe a procedure for serializing an RDF triples-set as a property table in time $\Theta(n)$ where $n$ is the number of triples. The procedure runs in two passes. In a first pass over the set of triples, the procedure would record the distinct properties and subjects in two respective sets, $S_1$ and $S_2$. The \emph{property schema} would then be built using the RDF dataset name and the \emph{field set} $S_1 \cup \{subject\}$. The property table would itself be initialized, with the \emph{subject} column populated using $S_2$. An \emph{index} to the subject column would also be built, with the subject as key and the row position as value. In the second pass over the triples, the cells of the table are incrementally updated with each encountered triple. An associative map admits an efficient implementation for the second pass.

For completeness, Algorithm \ref{rdf1} shows the inverse serialization; that is, how a property table is converted to a triples-set. Note that both the procedure above and Algorithm \ref{rdf1} are \emph{information-preserving}; no information is lost when we exchange representations.
 \begin{algorithm}
\caption{RDF Property Table to RDF Triples-set}
\begin{algorithmic}\label{rdf1}
\STATE \textbf{Input :} Property Table with property schema P(subject, $a_1, \ldots,a_n$) and $n$ properties
\STATE \textbf{Output :} Triples-set $P'$
\STATE \textbf{Method :}
\STATE{Initialize empty triples-set $P'$}
\FORALL{tuples $t$ in property table instance}
\STATE{Let $q=t(subject)$}
\FORALL{$a=a_1\ldots a_n$}
\IF{$t(a)$ is \emph{null}}
\STATE{continue}
\ENDIF
\STATE{Tokenize $t(a)$ using \emph{;} to obtain (possibly singleton) tokens-set $T'$}
\FORALL{tokens $t' \in T'$}
\STATE{Add triple $(q,a,t')$ to $P'$}
\ENDFOR
\ENDFOR
\ENDFOR
\STATE{Output $P'$}
\end{algorithmic}
\end{algorithm}  

Algorithm \ref{rdf1} is fairly simple and runs in worst-case time $O(mn)$, where $m$ is the number of subjects in the table, and $n$ is the total number of properties. We assume that each subject can be bound above by a constant number of object values (per property), a reasonable assumption in real-world cases. If not true, dataset characteristics need to be known before run-time can be bound. We omit a proof that the procedure always yields the same triples-set $P'$ from which the table was derived, if it was indeed derived from such a table using the two-pass algorithm earlier described. As we noted in Section \ref{proptable}, implemented triplestores often \emph{natively} store RDF datasets as \emph{physical} property tables. In a real-world implementation, it is possible to exploit this \emph{systems-level} advantage and use a provided triplestore API to access the physical table in its native form and use it logically \cite{propertytable}. We cited this earlier as a potential advantage in using the property table, and not devising a \emph{new} data structure for resolving RDF-tabular heterogeneity in ER.  

\subsection*{Optimal DNF Schemes}
We replicate the optimization condition first formally stated by Bilenko et al. \cite{bilenkoblocking}. 
Assume a (perfectly labeled) training set of \emph{duplicate pairs} D, and \emph{non-duplicate pairs} N. Let $f$ be a blocking scheme from the space $\mathcal{F}$ of all possible DNF blocking schemes that may be constructed from subsets of the set $H_c$ of SBPs and terms. $\mathcal{F}$ provably has cardinality $2^{|H_c|}$, since each DNF scheme is merely a \emph{positive} DNF formula $f$ constructed by treating $H_c$ as the \emph{atoms-set} (Definition \ref{dnfbs}).
The \emph{optimal} blocking scheme $f^* \in \mathcal{F}$ satisfies the following objective:  
\begin{equation}
f^*=argmin_{f} \sum\limits_{\{r,s\}\in N} f(r,s)
\end{equation}
s.t.\\
\begin{equation}\label{epsilon}
\sum\limits_{\{r,s\}\in D}f^*(r,s) \geq |D|\epsilon
\end{equation}
$D$, $N$ and $\epsilon$ were defined earlier in the paper. We also stated the meaning of this condition intuitively, which is that the scheme must cover at least a fraction $\epsilon$ of the set $D$ while minimizing coverage of non-duplicates in $N$. 
Bilenko et al. were the first to prove that this optimization problem is, in fact, NP-Hard by reducing from Red-Blue Set Covering (see \emph{Set Covering} Appendix section) \cite{bilenkoblocking}. We refer the reader to the original work for that proof.
Note that the condition is generic in that $r$ and $s$ do not have to be from structurally homogeneous datasets, and the proof of Bilenko et al. does not assume such a restriction either \cite{bilenkoblocking}. This was one reason we were able to extend their system and use it as an instantiated supervised baseline for the final pipeline module in Figure \ref{pipelinefig}(a). It also implies the natural result that the problem remains NP-Hard for heterogeneous datasets. 

\subsection*{Set Covering}
We provide a \emph{generic} description of the Weighted Set Covering (SC) problem, along with Chvatal's greedy algorithm, which, despite being relatively simple, continues to be the best-known polynomial-time approximation scheme (called PTAS). The weighted instance of SC assumes (as input) a \emph{universe set} $\mathcal{U}$ with $n$ elements and a \emph{family} of $m$ sets $S=\{S_1,\ldots , S_m\}$, where each $S_i$ is a subset of $\mathcal{U}$. Each set in the family is associated with a \emph{weight}, $w(S_i)$ for all $i$ from 1 to $m$. Additionally, the condition $\bigcup_i S_i=\mathcal{U}$ is assumed to hold. The SC problem is to find a subfamily $\mathcal{C} \subseteq S$ such that the summed weights of all sets in $\mathcal{C}$ are \emph{minimized} subject to the (mandatory) condition that $\bigcup_{c \in \mathcal{C}} c=\mathcal{U}$ (denoted as the \emph{covering} condition for the following discussion).  The decision version of this problem is known to be NP-Complete, and the optimization version, NP-Hard \cite{chvatal}. This is also true for the non-weighted SC, which reduces to a special case of W-SC with each set assigned equal (usually unit) weight.

Chvatal's greedy algorithm can be stated simply as follows. Initialize $\mathcal{C}$ to be the empty set. Iterate over $S$ till the covering condition is met. In each iteration, pick a (previously unpicked) set $S_i$ with \emph{maximum} score $|S_i|/w(S_i)$. In other words, we greedily pick the set in the family (breaking ties arbitrarily) that covers the most elements per unit weight. It is straightforward to observe that this algorithm is polynomial time; even a simple approach can run in time \emph{O(mn)} if the loop body is properly implemented. A linear-time algorithm along the same lines is possible if more advanced data structures are used. We do not go into details of how to optimize this algorithm; in most cases, an efficient off-the-shelf implementation can be adapted. 

In his seminal work (in which he proposed this algorithm), Chvatal also proved that the final (summed) weight of the approximate answer $\mathcal{C}$ is greater than the optimal answer $\mathcal{C}^*$ by a factor of (at most) $H(d)$, where, for any $x \in \mathbb{Z}^+$, the function $H(x)=\Sigma_{i} 1/i$, for all integers $i \in [1,x]$ \cite{chvatal}. Note that $H(x) \leq ln(x)+1$. Here, $d$ is simply the cardinality of the largest set in $S$. Chvatal's logarithmic approximation ratio remains the best-possible, even decades later \cite{setcoverguarantee}.  

Many variants of SC have been proposed over the decades; an important one is the Red-Blue Set Cover (RB-SC) \cite{approx}, which is closely related to the supervised method of Bilenko et al. \cite{bilenkoblocking}. RB-SC takes a universe set $\mathcal{U}$ as input, with $\mathcal{U}=R \cup B$, where $R$ is the set of \emph{red} elements and $B$, the set of \emph{blue} elements. Note that $R \cap B$ is empty. Again, we are given a family $S$ of subsets, but $S$ is only constrained to cover all of $B$, not necessarily the full universe set $\mathcal{U}$. RB-SC needs to locate a subfamily $\mathcal{C}$ such that all blue elements are covered but with the number of \emph{distinct} red elements covered, minimized. Sets in the family are not associated with weights; weighted generalizations of RB-SC are not relevant for this discussion.

RB-SC seems similar to the ordinary SC, but it is `harder' in an approximation sense. The best known approximation ratio for RB-SC, first proved by Peleg, is 2$\sqrt{|\mathcal{U}|log|B|}$ \cite{approx}. In that paper, he proposed an approximation algorithm that was adopted by Bilenko et al. \cite{bilenkoblocking}.

In Section \ref{unsup}, we explained how to reduce our specific problem to W-SC, by treating $H_c$ as the family of subsets $S$, given that each SBP or term (in $H_c$) covers \emph{tuple pairs}, and with all \emph{duplicate} tuple pairs together comprising the universe set $\mathcal{U}=D$. We used the non-duplicates set $N$ only to calculate weights. On the other hand, Bilenko et al. performed a more \emph{intuitive} reduction to RB-SC, by treating red elements as analogous to non-duplicates and blue elements to duplicates, in a given training set. This direct reduction comes with weak approximation bounds, however, as the discussion above shows. Empirically, we believe that better approximation results led to \emph{at par} PC results (in Experiments 1-2) for the proposed method on four DPs, compared to the supervised baseline, and to \emph{better} PC results (with high significance) on the \emph{largest} DP (4). 

\subsection*{General Blocking Predicates}
The set $G$ of General Blocking Predicates (GBPs) and the parameter $k$ are used (along with mappings $Q$ in the heterogeneous case presented in the paper, or the field set $A$ presented in previous work \cite{mayankblocking,knoblockblocking,bilenkoblocking,semisup}) to build the \emph{search space} of terms and SBPs, $H_c$. Intuitively, $G$ constitutes the \emph{feature space} of the algorithm and choosing an appropriate $G$ is an important empirical consideration. Experiment 2 in Section \ref{experiments}, and the described \emph{follow-up} experiment in the subsequent discussion, demonstrate this point. Specifically, the experiments shows that if $G$ is expressive enough, setting $k$ to 1 is adequate. Recall that, in this paper, we adapted the original set $G$, first proposed by Bilenko et al. and \emph{supplemented} it with phonetic functions found in an open-source package\footnote{org.apache.commons.codec.language}, to make $G$ more expressive. For completeness, we first provide a brief description of the original set. Note that all GBPs below are \emph{case-insensitive}.

{\bf (1)  Exact Match:} Returns \emph{True} if input strings exactly match. 

{\bf (2) ContainsCommonToken:} Returns \emph{True} if input strings share a common token, based on common delimiters (such as comma, whitespace and semicolon).

{\bf (3) ContainsCommonInteger:} Returns \emph{True} if input strings share at least one common integer token. If no token is an integer in a given input string, \emph{False} is returned by default.

{\bf (4) ContainsCommonOrOffByOneInteger:} Same as above, except integers may be off by one. Note that if \emph{True} is returned by \emph{ContainsCommonInteger}, \emph{True} is also returned for this GBP. This demonstrates that GBPs may be correlated. 

{\bf (5-7) ContainsTokenWithSameNFirstChars:} Returns \emph{True} if the input strings share at least one token with a common N-character prefix. Implemented with N=3,5,7 to yield three (correlated) GBPs. 

{\bf (8-10) ContainsTokenWithCommonNGram:} Returns \emph{True} if the input strings share a common length-N contiguous subsequence of \emph{tokens}. Implemented with N=2,4,6.

In total, these yield 10 GBPs. Although these GBPs have been found to work quite well in previous work, including the original work in which they were first proposed \cite{bilenkoblocking}, a rationale was never provided for why \emph{specifically} each of them were included. We briefly attempt to do so here, based on our experimental observations.

GBPs 1-2  are appropriate for strings that have high token overlap or for alphanumeric codes (in product databases, for example) that tend to match exactly and have high correlation with duplicate classification. GBPs 3-4 are more appropriate for phone numbers, zip codes, street numbers, social security numbers, dates of birth and other numeric quantities that commonly occur in databases. GBPs 5-7 are empirically robust to \emph{data representation} issues; for example, GBP 5 would return \emph{True} for two address strings that spell `Avenue' as \emph{Avenue} or \emph{Ave}. GBPs 8-10 are restrictive versions of GBP 2, and thus, highly \emph{discriminative}. They rarely return \emph{True}, but when they do, it indicates strongly that the input strings are derived from a duplicate pair.

Note that Bilenko et al. included 10 \emph{additional} GBPs that were based on TF-IDF and were appropriate for \emph{homogeneous} datasets \cite{bilenkoblocking}. In pilot experiments (and also the Dumas preliminary experiments; see Figure \ref{dumas1}), we obtained the unsurprising result that these TF-IDF features had \emph{negative} correlation with heterogeneous BSL performance. These were therefore not included in the $G$ used in this paper. 

In the Apache open-source package, nine phonetic functions are implemented and all were included in the supplemented $G$. These are respectively \emph{Caverphone1, Caverphone2, ColognePhonetic, DoubleMetaphone, MatchRatingApproachEncoder, Metaphone, NYSIIS, RefinedSoundex} and \emph{Soundex}. Christen provides a good description (and evaluation) of these phonetic encodings in his comprehensive text \cite{datamatching}. Perhaps the most important advantage of phonetic functions is that they are robust to spelling variations (especially in names) that the other GBPs cannot easily accommodate (e.g. Kathryn vs. Catherine).   

Finally, it is important to note that each of these GBPs is associated with an \emph{indexing function} (see Definition \ref{if}). Typically, the associated indexing functions simply extract some characters, tokens or integers and return the extracted elements in a set (GBPs 1-10); similarly, the associated phonetic indexing functions tokenize the string and return a set containing the appropriate phonetic encoding (e.g. Soundex) of each token. GBPs cannot be \emph{arbitrary} boolean functions. As an example, the boolean function $EditDistance<0.5$ might seem like a legitimate GBP\footnote{It takes two strings as input and returns \emph{True} if the Edit distance is less than 0.5.}, but it is not evident how to frame it as a \emph{set-intersection} condition on outputs of (some) indexing function, as the original GBP definition (Definition \ref{gbp}) formally requires.  
\balancecolumns

\end{document}